\documentstyle[12pt,fleqn]{article}
\begin{document}
\pagenumbering{arabic}
\newcommand{\vecvar}[1]{\mbox{\boldmath$#1$}}
\def\tsum{\mathop{{\rm T}_{\Sigma}}}
\newcommand{\myref}[1]{(\ref{#1})}
\def\Define{\mathop{\stackrel{\rm def}{=}}}
\def\Hchat{{{\hat H}_{\rm C}}}
\def\Hstilde{{{\tilde H}_{\rm S}}}
\def\Hc{{H_{\rm C}}}
\def\Hs{{H_{\rm S}}}
\def\Lcheck{{\check L}}
\def\Qcheck{{\check Q}}
\def\Lm{{L^{-}}}
\def\Lp{{L^{+}}}
\def\Lpm{{L^{\pm}}}
\def\Lmp{{L^{\mp}}}
\def\Lmhat{{{\hat L}^{-}}}
\def\Lphat{{{\hat L}^{+}}}
\def\phighat{{{\hat \phi}_{\rm g}}}
\def\varphighat{{{\hat \varphi}_{\rm g}}}
\def\psigtilde{{{\tilde \psi}_{\rm g}}}
\def\dalpha{\alpha^{\dagger}}
\def\Ltilde{{{\tilde L}}}
\def\Qtilde{{{\tilde Q}}}
\def\Lcal{{\tilde {\cal L}}}
\def\ledo{{\stackrel{\rm D}{\leq}}}
\def\lesdo{{\stackrel{\rm d}{\leq}}}
\def\Sym{\Bigr|_{\rm Sym}}
\def\bp{b^{+}}
\def\Mae{\mbox{\hspace*{-20pt}}}
\def\Cmae{\mbox{\hspace*{-10pt}}}
\def\Usr{\mbox{\hspace*{20pt}}}
\def\Cusr{\mbox{\hspace*{10pt}}}
\hspace*{\fill}{\sf UT/Wu016, cond-mat/9706156}
\begin{flushleft}

{\Large\bf The Calogero Model: Integrable Structure
and Orthogonal Basis}

\vspace{10pt}

{\Large Miki Wadati and Hideaki Ujino}

\vspace{10pt}

{\large \it Department of Physics, Graduate School of Science,

University of Tokyo,

Hongo 7--3--1, Bunkyo-ku, Tokyo 113, Japan}

\end{flushleft}\vspace{10pt}

\noindent{\bf Abstract.}
Integrability, algebraic structures and orthogonal basis of
the Calogero model are studied by the quantum Lax and Dunkl
operator formulations.
The commutator algebra among operators including conserved operators
and creation-annihilation operators has the structure of
the W-algebra. Through an algebraic construction of the simultaneous
eigenfunctions of all the commuting conserved operators, we show
that the Hi-Jack (hidden-Jack) polynomials, which are an multi-variable
generalization of the Hermite polynomials, form the orthogonal basis.

\renewcommand{\thefootnote}{}
\footnotetext{Based on the talk given by MW at the Workshop on the
Calogero-Moser-Sutherland models at CRM, Montreal, March, 1997.
To appear in the proceedings of the workshop.}

\section{Introduction}\hspace*{\parindent}
In memory of the pioneering works in 70's~\cite{Calogero_1,%
Moser_1,Sutherland_1,Sutherland_2},
a class of one-dimensional quantum many-body systems with inverse-square
long-range interactions are generally called the Calogero-Moser-Sutherland
models. The celebrated Hamiltonians are
\begin{eqnarray}
  & & \Mae \mbox{Calogero-Moser: } H_{\rm CM} =
  \frac{1}{2}\sum_{j=1}^{N}p_{j}^{2}
  +\frac{1}{2}\sum_{\stackrel{\scriptstyle j,k=1}{j\neq k}}^{N}
  \frac{a^{2}-\hbar a}{(x_{j}-x_{k})^{2}},
  \label{eqn:Calogero-Moser}\\
  & & \Mae \mbox{Calogero: } \Hchat = 
  \frac{1}{2}\sum_{j=1}^{N}\bigl(p_{j}^{2}+\omega^{2}x_{j}^{2}\bigr)
  +\frac{1}{2}\sum_{\stackrel{\scriptstyle j,k=1}{j\neq k}}^{N}
  \frac{a^{2}-\hbar a}{(x_{j}-x_{k})^{2}},
  \label{eqn:Calogero_model}\\
  & & \Mae \mbox{Sutherland: }
  \Hstilde=\frac{1}{2}\sum_{j=1}^{N}p_{j}^{2}
  +\frac{1}{2}\sum_{\stackrel{\scriptstyle j,k=1}{j\neq k}}^{N}
  \frac{a^{2}-\hbar a}{\sin^{2}(x_{j}-x_{k})},
  \label{eqn:Sutherland_model}
\end{eqnarray}
where the constants $N$, $a$ and $\omega$ are the particle number,
the coupling parameter and the strength of the external harmonic well,
respectively.
The momentum operator $p_{j}$ is given by a differential operator,
$p_{j} = -{\rm i}\hbar\frac{\partial}{\partial x_{j}}$.
The Calogero and Sutherland models are a harmonic confinement
and a periodic version of the Calogero-Moser model, respectively.
Thus these two models have discrete energy spectra, whereas the other
has continuous one.
From now on, we set the Planck constant at unity, $\hbar=1$.

The Lax formulation for the classical Calogero-Moser model was discovered
by Moser~\cite{Moser_1}. Let us introduce two $N\times N$ Hermitian matrices:
\begin{eqnarray*}
  & & \Mae L_{ij}=p_{i}\delta_{ij}+{\rm i}a(1-\delta_{ij})
  \frac{1}{x_{i}-x_{j}},
  \\
  & & \Mae M_{ij}= a\delta_{ij}
  \sum_{\stackrel{\scriptstyle l=1}{l\neq i}}^{N}
  \frac{1}{(x_{i}-x_{l})^{2}} -a(1-\delta_{ij})
  \frac{1}{(x_{i}-x_{j})^{2}}.
\end{eqnarray*}
We call them Lax pair.
The classical Calogero-Moser Hamiltonian is given by eq.
\myref{eqn:Calogero-Moser} with $p_{j}=\frac{{\rm d}x_{j}}{{\rm d}t}$
and $\hbar=0$.
The time evolution of the $L$-matrix is expressed
as the Lax equation,
\begin{equation}
  \frac{{\rm d}L}{{\rm d}t}
  = \bigl\{L,H_{\rm CM}^{\rm cl}\bigr\}_{\rm P}
  = \bigl[L,{\rm i}M\bigr],
  \label{eqn:CM_Lax_equation}
\end{equation}
where the Poisson bracket is defined
by $\{f,g\}_{\rm P}\Define\sum_{j}\bigl(\frac{\partial f}{\partial x_{j}}
\frac{\partial g}{\partial p_{j}}-\frac{\partial g}{\partial x_{j}}
\frac{\partial f}{\partial p_{j}}\bigr)$.
Thanks to the trace identity
for c-number-valued matrices, ${\rm Tr}AB={\rm Tr}BA$, we can easily
see that the trace of the power of the $L$-matrix,
$I_{n}^{\rm cl}\Define{\rm Tr}L^{n}$, gives the conserved quantities:
$\frac{{\rm d}I_{n}^{\rm cl}}{{\rm d}t}={\rm Tr}[L^{n},{\rm i}M]=0$.
Using the classical $r$-matrix~\cite{Avan_1}
or the generalized Lax equations for higher conserved
quantities \cite{Barucchi_1},
we can show that the conserved quantities are Poisson-commutative,
$\{I_{n}^{\rm cl},I_{m}^{\rm cl}\}_{P}=0$, which proves
the integrability of the classical Calogero-Moser model in Liouville's sense.
Natural quantization of the Lax equation for the classical case
\myref{eqn:CM_Lax_equation} by correspondence principle,
$\{\clubsuit,\spadesuit\}_{\rm P}\rightarrow{\rm -i}[\clubsuit,\spadesuit]$,
gives an equality for the quantum Calogero-Moser
Hamiltonian \myref{eqn:Calogero-Moser}.
However, the trace trick is not available to construct the commuting
conserved operators for the quantum case because of the non-commutativity
of the canonical conjugate variables. The initial motivation of our study
was to find out a way to construct the conserved operators for the quantum
models using the Lax formulation. The key of our idea is the sum-to-zero
condition of the $M$-matrix:
\[
  \sum_{j=1}^{N}M_{jk}=0,\;\;\sum_{j=1}^{N}M_{kj}=0,\;\;
  \mbox{for }k=1,2,\cdots,N.
\]
This property tells us that the (commuting) conserved operators can be
obtained by summing up all the matrix elements of the powers of the
$L$-matrix instead of taking traces \cite{Ujino_0,Ujino_1},
\[
  I_{n}^{\rm CM}=\sum_{j,k=1}^{N}(L^{n})_{jk}\Define\tsum L^{n}\Rightarrow
  [H_{\rm CM},I_{n}^{\rm CM}]=\tsum[L^{n},M]=0,
\]
which proves the quantum integrability, or the existence of sufficiently
many conserved operators, of the Calogero-Moser model.
Encouraged by the result, we further investigated the integrable
structure of the Calogero model \myref{eqn:Calogero_model} through
the quantum Lax formulation.

\section{Integrability and Algebraic Structure}\hspace*{\parindent}
Let us start from the Lax equation of the Calogero Hamiltonian
\myref{eqn:Calogero_model},
\[
  -{\rm i}\frac{{\rm d}\Lpm}{{\rm d}t} =
  \bigl[\hat{H}_{\rm C},L^{\pm}\bigr] = 
  \bigl[L^{\pm},M\bigr]\pm\omega L^{\pm},
\]
where the new matrices $\Lpm$ are defined by $\Lpm\Define L\pm Q$,
$Q_{jk}\Define{\rm i}x_{j}\delta_{jk}$.
Using the sum-to-zero trick, we can get the
conserved operators of the Calogero model as follows:
\begin{eqnarray*}
  & & \Mae \hat{I}_{n}\Define\tsum (\Lp\Lm)^{n},\;\;
  [\hat{H}_{\rm C},\hat{I}_{n}]=\tsum[(\Lp\Lm)^{n},M]=0,\nonumber\\
  & & \Mae \hat{I}_{1}=2\hat{H}_{\rm C}-N\omega\bigl(Na+(1-a)\bigr),\;\;
  \hat{I}_{n}=\sum_{j=1}^{N}p_{j}^{2n}+\cdots.
\end{eqnarray*}
Mutual commutativity of the above conserved operators is verified
rather easily by the Dunkl operator
formulation~\cite{Dunkl_1,Polychronakos_1}.
Introducing the coordinate exchange operator,
\[
  (K_{lk}f)(\cdots,x_{l},\cdots,x_{k},\cdots)
  = f(\cdots,x_{k},\cdots,x_{l},\cdots),
\]
we define the creation-annihilation like operators as
\begin{eqnarray}
  & & \Mae c_{l}^{\dagger} = p_{l}
  +{\rm i}a\sum_{\stackrel{\scriptstyle k=1}{k\neq l}}^{N}
  \frac{1}{x_{l}-x_{k}}K_{lk} + {\rm i}\omega x_{l},\nonumber\\
  & & \Mae c_{l} = p_{l}
  +{\rm i}a\sum_{\stackrel{\scriptstyle k=1}{k\neq l}}^{N}
  \frac{1}{x_{l}-x_{k}}K_{lk} - {\rm i}\omega x_{l}.
  \label{eqn:creation-annihilation-like_op}
\end{eqnarray}
Commutation relations among the creation-annihilation operators are
\begin{eqnarray*}
  & & \Mae \bigl[c_{l}^{\dagger},c_{m}^{\dagger}\bigr] = 0,\;\;
  \bigl[c_{l},c_{m}\bigr] = 0,\\
  & & \Mae \bigl[c_{l},c_{m}^{\dagger}\bigr] = 2\omega\delta_{lm}
  \bigl(1+a\sum_{\stackrel{\scriptstyle k\neq l}{k=1}}^{N}
  K_{lk}\bigr)-2\omega a(1-\delta_{lm})K_{lm},
\end{eqnarray*}
which prove that the Hermitian operators,
${\sf I}_{n}\Define\sum_{j=1}^{N}(c_{j}^{\dagger}c_{j})^{n}$,
are commuting operators.
We denote the restriction of the operand to the space of symmetric
functions by $\Sym$. Under the restriction, the conserved operators
$\hat{I}_{n}$ and commuting Hermitian operators ${\sf I}_{n}$
are considered to be the same, $\hat{I}_{n}\Sym={\sf I}_{n}\Sym$. Thus
we have proved the quantum integrability of the Calogero model.

We can recursively construct generalized Lax equations for a family
of operators $O_{m}^{p}$, $m,p=1,2,\cdots$,
which reveal the W-algebraic structure of the Calogero model
\cite{Ujino_2}.
The operators are defined by the sum of all the matrix elements
of the Weyl ordered product of $p$ $\Lp$s and $m$ $\Lm$s:
\begin{eqnarray*}
  & & \Mae 
  O^{p}_{m} \Define \tsum\bigl[(L^{+})^{p}(L^{-})^{m}\bigr]_{\rm W},
  \\
  & & \Mae \bigl[(L^{+})^{p}(L^{-})^{m}\bigr]_{\rm W}
  \Define \frac{p!m!}{(p+m)!}
  \sum_{\stackrel{\mbox{\scriptsize all possible}}{\mbox{\scriptsize order}}}
  (L^{+})^{p}(L^{-})^{m}.
\end{eqnarray*}
The generalized Lax equations for the operators $O_{m}^{p}$ are
\begin{eqnarray}
  \bigl[O_{m}^{p},L^{\pm}\bigr]
  & = & \bigl[L^{\pm},M_{m}^{p}\bigr]
  +m\omega(1\pm 1)\bigl[(L^{+})^{p}(L^{-})^{m-1}\bigr]_{\rm W}\nonumber\\
  & & -p\omega(1\mp 1)\bigl[(L^{+})^{p-1}(L^{-})^{m}\bigr]_{\rm W},
  \label{eqn:generalized_Lax}
\end{eqnarray}
and the $M_{m}^{p}$ matrices satisfy the sum-to-zero condition:
\[
  \sum_{j=1}^{N}(M_{m}^{p})_{jk} = 0, \;\;
  \sum_{j=1}^{N}(M_{m}^{p})_{kj}=0.
\]
The Hamiltonian $\hat{H}_{\rm C}$ belongs to the operator family,
$2\hat{H}_{\rm C}=O_{1}^{1}$. The operators $O_{n}^{n}$ are conserved
operators, though they do not commute each other.
The family has two interesting subsets of commuting non-Hermitian
operators,
\begin{equation}
  B_{n}^{\dagger}\Define O_{0}^{n},\;\;B_{n}\Define O_{n}^{0},
  \;\; n=1,2,\cdots,
  \label{eqn:power-sum_creation-annihilation_operator}
\end{equation}
which we call power-sum creation-annihilation operators. They
will play an important role in the algebraic construction of
the energy eigenfunctions in the next section.

Let us introduce operators $W_{n}^{(s)}$:
\[
  W_{n}^{(s)}\Define\frac{1}{4\omega}O_{s+n-1}^{s-n-1},\;\;
  s\geq |n|+1,
\]
where the indices $n$ and $s$ are integer or half odd integer,
and respectively correspond to the Laurent mode and the conformal spin.
The commutator among the operators above is
\[
  \bigl[W_{n}^{(s)},W_{m}^{(t)}\bigr]
  = \bigl(n(t-1)-m(s-1)\bigr)W_{n+m}^{(s+t-2)}
  +{\cal P}_{n,m}^{(s,t)}(W_{l}^{(u)}),
\]
where ${\cal P}_{n,m}^{(s,t)}(W_{l}^{(u)})$ is a polynomial
of $W_{l}^{(u)}$, $u\leq s+t-3$, $l\leq n+m$. The polynomial is generated
while the products of $\Lpm$-matrices are rearranged into the Weyl ordered
products by replacements of $\Lp$ and $\Lm$,
\[
  \bigl[L^{+},L^{-}\bigr]=2\omega \bigl((a-1){\bf 1}- aT\bigr),\;
  {\bf 1}_{jk}=\delta_{jk},\;T_{jk}=1.
\]
In terms of the W-operators,
conserved operators and power-sum creation-annihilation operators
are respectively expressed as $O_{n}^{n}\propto W_{0}^{n+1}$,
$B_{n}^{\dagger}\propto W_{-\frac{n}{2}}^{(\frac{n}{2}+1)}$ and
$B_{n}\propto W_{\frac{n}{2}}^{(\frac{n}{2}+1)}$.

For the classical case,
the W-algebraic structure of the Calogero model was discovered
by the collective field theory~\cite{Avan_2} and the
classical $r$-matrix method~\cite{Avan_5}. The quantum collective
field theory also possesses the W-algebraic
structure~\cite{Avan_3,Avan_4}, though its
relationship with the quantum Calogero model is not directly confirmed.
An $SU(\nu)$ generalization of our approach is presented in \cite{Ujino_3}.

\section{Perelomov Basis}\hspace*{\parindent}
The eigenvalue problem of the Calogero model was first solved by Calogero
\cite{Calogero_1}. Later, inspired by the simple form of its energy spectrum,
Perelomov tried an algebraic construction of the energy eigenfunctions%
~\cite{Perelomov_1}. In what follows,
we shall complete the Perelomov's approach~\cite{Ujino_2}.
The generalized Lax equations \myref{eqn:generalized_Lax}
for the power-sum creation operators
\myref{eqn:power-sum_creation-annihilation_operator}
yield the following commutators:
\begin{equation}
  \bigl[\hat{H}_{\rm C},B_{n}^{\dagger}\bigr]=n\omega B_{n}^{\dagger},\;\;
  \bigr[B_{n}^{\dagger},B_{m}^{\dagger}\bigr]=0,\;\;n,m=1,2,\cdots,N.
  \label{eqn:Perelomov_operators}
\end{equation}
To construct all the eigenfunctions for the $N$-body Calogero model,
we need $N$ creation operators. By straightforward calculations of
the commutators \myref{eqn:Perelomov_operators},
Perelomov presented three creation operators, $B_{n}^{\dagger}$,
$n=2,3,4$. The quantum Lax formulation provides
an easy way to make sufficient number of such operators.
By successive operations of the power-sum creation operators on
the ground state wave function, we can get all the excited states.
The Calogero Hamiltonian is cast into the following form,
\[
  \hat{H}_{\rm C} = \frac{1}{2}\tsum\Lp\Lm +\frac{1}{2}N\omega
  \bigl(Na+(1-a)\bigr)
  = \frac{1}{2}\sum_{j=1}^{N}h_{j}^{\dagger}h_{j}+E_{\rm g},
\]
where the operators $h_{j}^{\dagger}$ and $h_{j}$ are defined as
\begin{eqnarray*}
   & & \Mae h_{j}^{\dagger}\Define\sum_{k=1}^{N}L^{+}_{kj}
   =p_{j}+{\rm i}\omega x_{j}-{\rm i}a
   \sum_{\stackrel{\scriptstyle k=1}{k\neq j}}^{N}\frac{1}{x_{j}-x_{k}},\\
   & & \Mae h_{j}\Define\sum_{k=1}^{N}L^{-}_{jk}
   =p_{j}-{\rm i}\omega x_{j}+{\rm i}a
   \sum_{\stackrel{\scriptstyle k=1}{k\neq j}}^{N}\frac{1}{x_{j}-x_{k}}.
\end{eqnarray*}
Thus the differential equations, $h_{j}|0\rangle=0$, $j=1,2,\cdots,N$,
are the sufficient conditions for the ground state $|0\rangle$.
In the coordinate representation,
the solution is expressed as the (real) Laughlin wave function:
\begin{equation}
   \langle x|0\rangle=\prod_{1\leq j<k\leq N}|x_{j}-x_{k}|^{a}
   \exp(-\frac{1}{2}\omega\sum_{j=1}^{N} x_{j}^{2}).
  \label{eqn:ground_state}
\end{equation}

As is similar to the free boson case,
the excited states are labeled by the Young diagram,
\[
  \lambda=
  \{\lambda_{1}\geq\lambda_{2}\geq\cdots\geq\lambda_{N}\geq 0\}
  \in Y_{N},
\]
where
$\lambda_{k}$, $k=1,2,\cdots,N$, are nonnegative integers.
Conventionally, we omit zeroes and use superscript for a sequence
of the same numbers, e.g., $\{4,1^{2}\}=\{4,1,1,0,\cdots,0\}$.
The excited state labeled by the Young diagram $\lambda$ is given by
  \begin{eqnarray}
    & & \Mae |\lambda\rangle=\prod_{k=1}^{N}
    (B_{k}^{\dagger})^{\lambda_{k}-\lambda_{k+1}}|0\rangle,
    \;\; \lambda_{N+1}=0,\nonumber\\
    & & \Mae \hat{H}_{\rm C}=(|\lambda|\omega+E_{\rm g})|\lambda\rangle
    \Define E(\lambda)|\lambda\rangle,
    \label{eqn:Perelomov_construction}
  \end{eqnarray}
where $|\lambda |$ denotes the weight of the Young diagram,
$|\lambda |\Define\sum_{k=1}^{N}\lambda_{k}$.
The above energy spectrum has the same form as that of non-interacting
bosons confined in an external harmonic well up to the ground state
energy. In other words, the inverse-square interactions just shift
the ground state energy. The multiplicity of the $n$-th energy level,
$n\omega +E_{\rm g}$, is equal to the number of the Young diagrams of
the weight $n$,
$\#\{\lambda ||\lambda|=n\}$. This is also the same as that of the
non-interacting case. Thus we have algebraically constructed a basis of the
eigenfunctions of the Calogero Hamiltonian.

\section{Diagonalization of $\hat{I}_{2}$}\hspace*{\parindent}
The algebraic construction {\it \`{a} la} Perelomov
generates a basis of the eigenfunctions of the
Calogero Hamiltonian. Unfortunately, the basis is not orthogonal.
To make an orthogonal basis from a basis, we usually try the Gram-Schmidt
method. Here we take another way. As we have confirmed before, the
Calogero model has a set of commuting conserved operators, which means
existence of simultaneous eigenfunctions for them. The simultaneous
eigenfunctions should be the orthogonal basis because they must be
non-degenerate eigenfunctions of Hermitian operators. As the first step
of our approach, we get some simultaneous eigenfunctions of the Hamiltonian
and the second conserved operator $\hat{I}_{2}$, and observe their
properties~\cite{Ujino_4}.

Since the Hamiltonian and $\hat{I}_{2}$ commute,
the matrix representation of $\hat{I}_{2}$ on the Perelomov basis
has a block-diagonalized form and each block consists of the
wave functions of a weight (energy eigenvalue).
By a straightforward calculation of commutators between $\hat{I}_{2}$
and $B_{n}^{\dagger}$, we calculated the first seven blocks,
whose weights are from zero to six, of
the matrix representation and their eigenvalues. The eigenvalues
imply the general form of the eigenvalue of $\hat{I}_{2}$:
\[
  \hat{E}_{2}(\lambda)=4\omega^{2}\sum_{k=1}^{N}\bigl(
  (\lambda_{k})^{2}+a(N+1-2k)\lambda_{k}\bigr).
\]
Though a combination of $E(\lambda)$ and $\hat{E}_{2}(\lambda)$ removes most of
the degeneracies, there still remain degeneracies for the states
whose weights larger than or equal to six. For example, the following
two pairs of the Young diagrams with the weight six give such degeneracies:
\begin{eqnarray*}
  \bigl\{4,1^{2}\bigr\},\; \bigl\{3^{2}\bigr\} & \rightarrow &
  \hat{E}_{2}=4\omega^{2}\bigl(18+6a(N-2)\bigr),\\
  \bigl\{3,1^{3}\bigr\},\; \bigl\{2^{3}\bigr\} & \rightarrow &
  \hat{E}_{2}=4\omega^{2}\bigl(12+6a(N-3)\bigr).
\end{eqnarray*}
It is interesting that the pairs have a common property.
We can not compare two Young diagrams of each pair by
the dominance order. The dominance order $\ledo$, sometimes
called the natural partial order, is defined as follows:
\[
  \mu\ledo\lambda\Leftrightarrow |\mu|=|\lambda| 
  \mbox{ and }
  \sum_{k=1}^{l}\mu_{k}\leq\sum_{k=1}^{l}\lambda_{k}
  \mbox{ for all }l.
\]
We can readily confirm that the Young diagrams of each pair are
incomparable in the dominance order,
\begin{eqnarray*}
  & & \Mae \bigl\{4,1^{2}\bigr\}\stackrel{\rm D}{\not\leq}\bigl\{3^{2}\bigr\}
  \mbox{ and }
  \bigl\{3^{2}\bigr\}\stackrel{\rm D}{\not\leq}\bigl\{4,1^{2}\bigr\},\\
  & & \Mae \bigl\{3,1^{3}\bigr\}\stackrel{\rm D}{\not\leq}\bigl\{2^{3}\bigr\}
  \mbox{ and }
  \bigl\{2^{3}\bigr\}\stackrel{\rm D}{\not\leq}\bigl\{3,1^{3}\bigr\}.
\end{eqnarray*}
The specific observation above is, in fact, a general fact.
We cannot define the dominance order between any pair of distinct
Young diagrams $\lambda$ and $\mu$ of a weight that share the common
eigenvalue $\hat{E}_{2}$~\cite{Stanley_1,Ujino_5}, i.e.,
\[
  |\lambda|=|\mu| \mbox{ and } \hat{E}_{2}(\lambda)=\hat{E}_{2}(\mu)
  \Rightarrow \lambda\stackrel{\rm D}{\not\leq}\mu \mbox{ and }
  \mu\stackrel{\rm D}{\not\leq}\lambda.
\]

We calculated the eigenvectors of the blocks with weights up to three
in the matrix representation of $\hat{I}_{2}$. The eigenvectors correspond
to seven simultaneous eigenfunctions of $\hat{H}_{\rm C}$ and $\hat{I}_{2}$.
Since the eigenvalues $E$ and $\hat{E}_{2}$ for the seven functions have no
degeneracy, they belong to the orthogonal basis and also to the simultaneous
eigenfunctions of all the commuting conserved operators $\hat{I}_{n}$
of the Calogero model. The eigenfunction of the Calogero model is factorized
into the ground state wave function \myref{eqn:ground_state}
and a symmetric polynomial. Symmetric polynomial parts of the seven
simultaneous eigenfunctions, which we denote by $[\lambda]$, are
\begin{eqnarray*}
  & & \Mae \bigl[0\bigr] = 1,\;\;\bigl[1\bigr] = m_{1},
  \;\;\bigl[1^{2}\bigr] =
  m_{1^{2}}+\frac{a}{2\omega}\frac{N(N-1)}{2},\\
  & & \Mae \bigl[2\bigr] =
  (1+a)m_{2}+2am_{1^{2}}
  -\frac{1}{2\omega}N(Na+1),\\
  & & \Mae \bigl[1^{3}\bigr] =m_{1^{3}}
  +\frac{1}{2\omega}a\frac{(N-1)(N-2)}{2}m_{1},\\
  & & \Mae \bigl[2,1\bigr] = 
  (2a+1)m_{2,1}+6am_{1^{3}}
  -\frac{1}{2\omega}(1-a)(N-1)(Na+1)m_{1},\nonumber\\
  & & \Mae \bigl[3\bigr] = 
  (a^{2}+3a+2)m_{3}+3a(a+1)m_{2,1}+6am_{1^{3}}\nonumber\\
  & & \Cusr -\frac{3}{2\omega}(a^{2}N^{2}+3aN+2)m_{1},
\end{eqnarray*}
where $m_{\lambda}$ is the monomial symmetric polynomial defined by
\[
  m_{\lambda}(x_{1},\cdots,x_{N})=
  \sum_{\stackrel{\scriptstyle \sigma\in S_{N},\;{\rm distinct}}
  {\rm permutations}}(x_{\sigma(1)})^{\lambda_{1}}\cdots
  (x_{\sigma(N)})^{\lambda_{N}}.
\]
Note that the sum over $S_{N}$ is performed so that any monomial
in the summand appears only once. In the above expressions, we notice
that all the seven symmetric polynomials share a common property,
triangularity. Namely,
the seven polynomials $[\lambda]$ are expanded by the monomial symmetric
polynomials $m_{\mu}$ whose Young diagram $\mu$ is smaller than or equal
to the Young diagram $\lambda$ in the weak dominance order $\lesdo$, i.e.,
\[
  \mu\lesdo\lambda\Leftrightarrow
  \sum_{k=1}^{l}\mu_{k}\leq\sum_{k=1}^{l}\lambda_{k}
  \mbox{ for all $l$}.
\]
The observation means that we can uniquely identify
the simultaneous eigenfunctions of the first two conserved
operators of the Calogero model just by the first two eigenvalues and
triangularity up to normalization.
We shall confirm the existence of such functions by algebraically constructing
them.

\section{Hi-Jack Polynomials}\hspace*{\parindent}
Since our interest now concentrates on the symmetric polynomial parts of the
simultaneous eigenfunctions, we modify some operators to make them
suitable for the aim. A gauge transformation of the
creation-annihilation-like operators
\myref{eqn:creation-annihilation-like_op} yields the following
Dunkl operators:
\begin{eqnarray}
  \dalpha_{l} & \Define & \langle x|0\rangle
  \Bigl(-\frac{\rm i}{2\omega}\Bigr)c_{l}^{\dagger}
  \frac{1}{\langle x|0\rangle} \nonumber\\
  & = & -\frac{\rm i}{2\omega}\bigl(p_{l}
  +{\rm i}a\sum_{\stackrel{\scriptstyle k=1}{k\neq l}}^{N}
  \frac{1}{x_{l}-x_{k}}(K_{lk}-1) +2{\rm i}\omega x_{l}\bigr),
  \nonumber \\
  \alpha_{l} & \Define & \langle x|0\rangle {\rm i}c_{l}
  \frac{1}{\langle x|0\rangle} = {\rm i}\bigl(p_{l}
  +{\rm i}a\sum_{\stackrel{\scriptstyle k=1}{k\neq l}}^{N}
  \frac{1}{x_{l}-x_{k}}(K_{lk}-1)\bigr),
  \nonumber \\
  d_{l} & \Define & \dalpha_{l}\alpha_{l}.
  \label{eqn:Calogero_Dunkl}
\end{eqnarray}
The gauge transformation above removes the action on the ground state
wave function from the operators. Note that the definition of Hermiticity
of such gauge-transformed operators is modified and different
from the ordinary one.
Using the $d_{l}$-operators,
we define the normalized conserved operators:
\begin{eqnarray}
  & & \Mae I_{n}\Define\sum_{l=1}^{N}(d_{l})^{n}\Sym
  =\Bigl(\frac{1}{2\omega}\Bigr)^{n}\langle x|0\rangle \hat{I}_{n}
  \frac{1}{\langle x|0\rangle}\Sym,\nonumber\\
  & & \Mae \langle x|0\rangle \hat{H}_{\rm C}
  \frac{1}{\langle x|0\rangle}=\omega I_{1}+E_{\rm g}.
  \label{eqn:Calogero_conserved_operators_Dunkl}
\end{eqnarray}
We note that the Dunkl operators \myref{eqn:Calogero_Dunkl}
reduce to those for
the Sutherland model \myref{eqn:Sutherland_model} in the limit,
$\omega\rightarrow\infty$:
\begin{eqnarray}
  & & \Mae \dalpha_{l}\rightarrow z_{l},
  \nonumber\\
  & & \Mae \alpha_{l}\rightarrow\nabla_{l} = {\rm i}\bigl(p_{z_{l}}
  +{\rm i}a\sum_{\stackrel{\scriptstyle k=1}{k\neq l}}^{N}
  \frac{1}{z_{l}-z_{k}}(K_{lk}-1)\bigr),\;\;p_{z_{l}}\Define
  -{\rm i}\frac{\partial}{\partial z_{l}},
  \nonumber \\
  & & \Mae d_{l}\rightarrow D_{l} = z_{l}\nabla_{l}.
  \label{eqn:Sutherland_Dunkl}
\end{eqnarray}
We change the variables by
\[
  \exp{2{\rm i}x_{j}}=z_{j},\;\;\; j=1,2,\cdots,N,
\]
and denote the ground state wave function
and the ground state energy of the Sutherland model by
\begin{eqnarray*}
  \psigtilde & = & \prod_{1\leq j<k\leq N}|z_{j}-z_{k}|^{a}
  \prod_{j=1}^{N}z_{j}^{-\frac{1}{2}a(N-1)},
  \\
  \epsilon_{\rm g} & = & \frac{1}{6}a^{2}N(N-1)(N+1).
\end{eqnarray*}
Then the Sutherland Hamiltonian \myref{eqn:Sutherland_model} is
gauge-transformed to and related with the $D$-operator by
\begin{eqnarray*}
  \Hs-\epsilon_{\rm g}
  & = & \psigtilde^{-1}(\Hstilde-\epsilon_{\rm g})\psigtilde
  \nonumber\\
  & = & -2\sum_{j=1}^{N}(z_{j}p_{z_{j}})^{2} + {\rm i}a
  \sum_{\stackrel{\scriptstyle j,k=1}{j\neq k}}^{N}
  \frac{z_{j}+z_{k}}{z_{j}-z_{k}}(z_{j}p_{z_{j}}-z_{k}p_{z_{k}})
  \nonumber\\
  & = & 2\sum_{l=1}^{N}(D_{l})^{2}\Sym.
\end{eqnarray*}
Commutation relations among the Dunkl operators \myref{eqn:Calogero_Dunkl}
and the action of $\alpha_{l}$ on $1$ are
\begin{eqnarray}
  & & \Mae [\alpha_{l},\alpha_{m}] = 0,\;\;
  [\dalpha_{l},\dalpha_{m}] = 0,\nonumber\\
  & & \Mae[\alpha_{l},\dalpha_{m}]
  = \delta_{lm}
  \bigl(1+a\sum_{\stackrel{\scriptstyle k=1}{k\neq l}}^{N}K_{lk}\bigr)
  -a(1-\delta_{lm})K_{lm},\nonumber\\
  & & \Mae [d_{l},d_{m}] = a(d_{m}-d_{l})K_{lm},\;\;
  \alpha_{l} \cdot 1 = 0.
  \label{eqn:Needed_algebra_harmonic_Dunkl}
\end{eqnarray}
We should remark that the above relations do not explicitly depend on
the parameter $\omega$, which implies the Dunkl operators for the Sutherland
model \myref{eqn:Sutherland_Dunkl} also satisfy the above relations.
Hence the Calogero and Sutherland models
share the same algebraic structure~\cite{Ujino_6,Ujino_7,Ujino_8}.
To put it another way, the theory of the Calogero model is a one-parameter
deformation of that of the Sutherland model.
Thus the simultaneous eigenfunction of the Calogero model is expected to be
a one-parameter deformation of that of the Sutherland model, which is
known to be the Jack polynomial~\cite{Jack_1}.
In the following, we call the simultaneous eigenfunction of the Calogero
model Hi-Jack (hidden-Jack) polynomial.

Using the normalized conserved operators
\myref{eqn:Calogero_conserved_operators_Dunkl}, we define the Hi-Jack
polynomials $j_{\lambda}(\vecvar{x};\omega,1/a)$
in a similar fashion to a definition of the Jack polynomials:
\begin{eqnarray}
  & & \Mae I_{1}j_{\lambda}(\vecvar{x};\omega,1/a) =
  \sum_{k=1}^{N}\lambda_{k}j_{\lambda}(\vecvar{x};\omega,1/a)
  \Define E_{1}(\lambda)j_{\lambda}(\vecvar{x};\omega,1/a),
  \label{eqn:Hi-Jack_eigenfunction_1}\\
  & & \Mae I_{2}j_{\lambda}(\vecvar{x};\omega,1/a)
  = \sum_{k=1}^{N}
  \bigl(\lambda_{k}^{2}+a(N+1-2k)\lambda_{k}\bigr)
  j_{\lambda}(\vecvar{x};\omega,1/a)\nonumber\\
  & & \Define E_{2}(\lambda)j_{\lambda}(\vecvar{x};\omega,1/a),
  \label{eqn:Hi-Jack_eigenfunction_2}\\
  & & \Mae j_{\lambda}(\vecvar{x};\omega,1/a)
  = \sum_{\mu\lesdo\lambda}
  w_{\lambda\mu}(a,1/2\omega)
  m_{\mu}(\vecvar{x}),
  \label{eqn:Hi-Jack_triangularity}\\
  & & \Mae w_{\lambda\lambda}(a,\omega) = 1.
  \label{eqn:Hi-Jack_normalization}
\end{eqnarray}

We can prove the existence of the Hi-Jack polynomials
by explicit construction.
Following the results by Lapointe and Vinet on the Jack
polynomials~\cite{Lapointe_1}, we introduce the raising operators for the
Hi-Jack polynomials,
\begin{eqnarray*}
  & & \Mae b^{+}_{k} =
  \sum_{\stackrel{\scriptstyle J\subseteq \{1,2,\cdots,N\}}{|J|=k}}
  \dalpha_{J}d_{1,J},\;\mbox{for }
  k=1,2,\cdots,N-1,
  \\
  & & \Mae b^{+}_{N} =
  \dalpha_{1}\dalpha_{2}\cdots\dalpha_{N}.
\end{eqnarray*}
The operators, $\dalpha_{J}$ and $d_{1,J}$, stand for
\begin{eqnarray*}
  \dalpha_{J} & = & \prod_{j\in J}\dalpha_{j},
  \\
  d_{1,J} & = & (d_{j_{1}} + a)(d_{j_{2}} + 2a) 
  \cdots(d_{j_{k}} + ka),
\end{eqnarray*}
where $J$ is a subset of a set $\{1,2,\cdots,N\}$
whose number of elements $|J|$ is
equal to $k$, $J\subseteq\{1,2,\cdots,N\}$, $|J|=k$. 
From eq. \myref{eqn:Needed_algebra_harmonic_Dunkl}, 
we can verify an identity,
\begin{equation}
  (d_{i}+ma)(d_{j}+(m+1)a)\Sym^{\{i,j\}}
  = (d_{j}+ma)(d_{i}+(m+1)a)\Sym^{\{i,j\}},
  \label{eqn:Identity_di_and_dj}
\end{equation}
where $m$ is some integer. The symbol $\Sym^{J}$ where $J$ 
is some set of
integers means that the operands are restricted to the space which
is symmetric with respect to the exchanges of any indices
in the set $J$. 
This identity \myref{eqn:Identity_di_and_dj}
guarantees that the operator $d_{1,J}$ does not depend on the order
of the elements of a set $J$ when it acts on symmetric functions
and hence operation of the raising operators on symmetric functions
yields symmetric functions.
The function generated by the following Rodrigues formula,
\[
  j_{\lambda}(\vecvar{x};\omega,1/a) =
  C_{\lambda}^{-1}(b_{N}^{+})^{\lambda_{N}}
  (b_{N-1}^{+})^{\lambda_{N-1}-\lambda_{N}}
  \cdots (b_{1}^{+})^{\lambda_{1}-\lambda_{2}}\cdot 1,
\]
with the normalization constant $C_{\lambda}$ given by
  \[
    C_{\lambda}=
    \prod_{k=1}^{N-1}C_{k}(\lambda_{1},\lambda_{2},\cdots,\lambda_{k+1};a),
  \]
  where
  \begin{eqnarray*}
    C_{k}(\lambda_{1},\lambda_{2},\cdots,\lambda_{k+1};a)
    & = & (a)_{\lambda_{k}-\lambda_{k+1}}
    (2a+\lambda_{k-1}-\lambda_{k})_{\lambda_{k}-\lambda_{k+1}}
    \nonumber\\
    & & \cdots(ka+\lambda_{1}-\lambda_{k})_{\lambda_{k}-\lambda_{k+1}},
  \end{eqnarray*}
satisfies the definition of the Hi-Jack polynomial
$j_{\lambda}(\vecvar{x};\omega,1/a)$.
The symbol $(\beta)_{n}$ in the above expression is the
Pochhammer symbol, that is, $(\beta)_{n}=\beta(\beta+1)\cdots(\beta+n-1)$,
$(\beta)_{0}\Define 1$.

The first seven Hi-Jack polynomials are, for instance, given as
follows:
\begin{eqnarray*}
  & & \Mae j_{0}(\vecvar{x};\omega,1/a) = J_{0}(\vecvar{x};1/a)
  =m_{0}(\vecvar{x})=1,\\
  & & \Mae j_{1}(\vecvar{x};\omega,1/a) = J_{1}(\vecvar{x};1/a)
  =m_{1}(\vecvar{x}),\\
  & & \Mae j_{1^{2}}(\vecvar{x};\omega,1/a) = J_{1^{2}}(\vecvar{x};1/a)
  +\frac{a}{2\omega}\frac{N(N-1)}{2}J_{0}(\vecvar{x};1/a)\nonumber\\
  & & = m_{1^{2}}(\vecvar{x})
  +\frac{a}{2\omega}\frac{N(N-1)}{2}m_{0}(\vecvar{x}),\\
  & & \Mae (a+1)j_{2}(\vecvar{x};\omega,1/a) = (a+1)J_{2}(\vecvar{x};1/a)
  -\frac{1}{2\omega}N(Na+1)J_{0}(\vecvar{x};1/a)\nonumber\\
  & & = (a+1)m_{2}(\vecvar{x})+2am_{1^{2}}(\vecvar{x})
  -\frac{1}{2\omega}N(Na+1)m_{0}(\vecvar{x}),\\
  & & \Mae j_{1^{3}}(\vecvar{x};\omega,1/a) = J_{1^{3}}(\vecvar{x};1/a)
  +\frac{1}{2\omega}a\frac{(N-1)(N-2)}{2}J_{1}(\vecvar{x};1/a)
  \nonumber\\
  & & =  m_{1^{3}}(\vecvar{x})
  +\frac{1}{2\omega}a\frac{(N-1)(N-2)}{2}m_{1}(\vecvar{x}),\\
  & & \Mae (2a+1)j_{2,1}(\vecvar{x};\omega,1/a) \nonumber\\
  & & = (2a+1)J_{2,1}(\vecvar{x};1/a)
  -\frac{1}{2\omega}(1-a)(N-1)(Na+1)J_{1}(\vecvar{x};1/a)\nonumber\\
  & & = (2a+1)m_{2,1}(\vecvar{x})+6am_{1^{3}}(\vecvar{x})\nonumber\\
  & & \Cusr -\frac{1}{2\omega}(1-a)(N-1)(Na+1)m_{1}(\vecvar{x}),
  \\
  & & \Mae (a^{2}+3a+2)j_{3}(\vecvar{x};\omega,1/a) \nonumber\\
  & & = (a^{2}+3a+2)J_{3}(\vecvar{x};1/a)
  -\frac{3}{2\omega}(a^{2}N^{2}+3aN+2)J_{1}(\vecvar{x};1/a)
  \nonumber\\
  & & = (a^{2}+3a+2)m_{3}(\vecvar{x})+3a(a+1)m_{2,1}(\vecvar{x})
  +6a^{2}m_{1^{3}}(\vecvar{x})\nonumber\\
  & & \Cusr -\frac{3}{2\omega}(a^{2}N^{2}+3aN+2)m_{1}(\vecvar{x}),
\end{eqnarray*}
where the symbol $J_{\lambda}(\vecvar{x};1/a)$ denotes the Jack polynomial.
The explicit forms also show the fact that the Hi-Jack polynomial reduces
to the Jack polynomial in the limit, $\omega\rightarrow\infty$,
\begin{equation}
  j_{\lambda}(\vecvar{x};\omega=\infty,1/a)=J_{\lambda}(\vecvar{x};1/a).
  \label{eqn:Hi-Jack_to_Jack}
\end{equation}
Besides the above relation, we have some other relations between
the Hi-Jack and Jack polynomials~\cite{Ujino_7,Ujino_8,Ujino_9}.
While the Hi-Jack polynomial is a one-parameter deformation
of the Jack polynomial, we can get the Hi-Jack polynomial
from the Jack polynomial by the following formula,
\[
  J_{\lambda}(\dalpha_{1},\dalpha_{2},\cdots,\dalpha_{N};1/a)\cdot 1
  = j_{\lambda}(\vecvar{x};\omega,1/a),
\]
which gives another relation 
between the Jack polynomials and the Hi-Jack polynomials.
In the above expansion, we have an observation, and it is generally
true, that increasing the order of $1/2\omega$ by one causes decreasing
of the weight of the symmetrized monomial by two. The fact yields an
stronger form of the triangularity:
\begin{eqnarray}
  & & \Mae j_{\lambda}(\vecvar{x};\omega,1/a)=
  \sum_{\stackrel{\scriptstyle \mu\lesdo\lambda\;{\rm and}}
  {|\mu|\equiv|\lambda|\;({\rm mod}2)}}
  \Bigl(\frac{1}{2\omega}\Bigr)^{(|\lambda|-|\mu|)/2}
  w_{\lambda\mu}(a)m_{\mu}(\vecvar{x}),\nonumber\\
  & & \Mae w_{\lambda\lambda}(a)=1.
  \label{eqn:strong_triangularity}
\end{eqnarray}
Combining eqs. \myref{eqn:Hi-Jack_to_Jack} and
\myref{eqn:strong_triangularity}, we have the following expansion
form of the Hi-Jack polynomial with respect to the Jack polynomial:
\begin{eqnarray*}
  & & \Mae j_{\lambda}(\vecvar{x};\omega,1/a) 
  \nonumber\\
  & & \Mae\; = J_{\lambda}(\vecvar{x};\omega,1/a)
  +\sum_{\stackrel{\scriptstyle \mu\lesdo\lambda\;
  {\rm and}\;|\mu|<|\lambda|}
  {{\rm and}\;|\mu|\equiv|\lambda|\;({\rm mod}2)}}
  \bigl(\frac{1}{2\omega}
  \bigr)^{(|\lambda|-|\mu|)/2}{\sf w}_{\lambda\mu}(a)
  J_{\mu}(\vecvar{x};1/a).
\end{eqnarray*}

Relationship between the Hi-Jack polynomials and the Perelomov basis
\myref{eqn:Perelomov_construction} is given as follows.
The power-sum creation operator $B_{k}^{+}$ is cast into the power-sum of
$\dalpha_{l}$-operators,
\[
  B_{k}^{\dagger}=(2{\rm i}\omega)^{k}
  \sum_{l=1}^{N}(\dalpha_{l})^{k}\Sym\Define
  (2{\rm i}\omega)^{k}{\sf p}_{k}(\vecvar{\dalpha})\Sym,
\]
and the Perelomov basis is expressed by the power-sum of the
$\dalpha_{l}$-operators as
\[
  \frac{\langle\vecvar{x}|\lambda\rangle}{\langle\vecvar{x}|0\rangle}
  =(2{\rm i}\omega)^{|\lambda|}
  \prod_{k=1}^{N}({\sf p}_{k}
  (\vecvar{\dalpha}))^{\lambda_{k}-\lambda_{k+1}}\cdot 1
  =(2{\rm i}\omega)^{|\lambda|}
  {\sf p}_{\lambda}(\vecvar{\dalpha})\cdot 1.
\]
Thus the transition matrix between the power-sums and Jack polynomials
$M(J,{\sf p})$,
\[
  J_{\lambda}(\vecvar{x},1/a)=
  \sum_{\stackrel{\scriptstyle \mu}{|\mu|=|\lambda|}}
  M(J,{\sf p})_{\lambda\mu}{\sf p}_{\mu}(\vecvar{x}),
\]
gives a relation between the Hi-Jack polynomials and the Perelomov basis,
\[
  j_{\lambda}(\vecvar{x};\omega,1/a)=(2{\rm i}\omega)^{-|\lambda|}
  \sum_{\stackrel{\scriptstyle \mu}{|\mu|=|\lambda|}}
  M(J,{\sf p})_{\lambda\mu}
  \frac{\langle\vecvar{x}|\mu\rangle}{\langle\vecvar{x}|0\rangle}.
\]

We have introduced the Hi-Jack polynomials as the simultaneous
eigenfunctions for the first two commuting conserved operators
with the triangularity.
As we shall see shortly,
they are non-degenerate simultaneous eigenfunctions
for all the commuting conserved operators of the Calogero model.
From a calculation of the action of $d_{l}$ operator on
a symmetrized monomial of $\dalpha_{k}$'s, $m_{\lambda}(\dalpha_{1},
\cdots,\dalpha_{N})$, we can prove the following expression:
\begin{equation}
  I_{n} j_{\lambda}(\vecvar{x};\omega,1/a)=
  \sum_{\stackrel{\scriptstyle \mu\ledo\lambda}
  {\mbox{\scriptsize or }|\mu|<|\lambda|}}
  w^{\prime}_{\lambda,\mu}(a,1/2\omega)m_{\mu}(\vecvar{x}).
  \label{eqn:Keep_triangularity}
\end{equation}
This means that operation of the conserved operators on the Hi-Jack
polynomials keeps their triangularity. Since the $n$-th conserved
operator commutes with the first and second conserved operators,
$[I_{1},I_{n}]=[I_{2},I_{n}]=0$, we can easily verify,
  \begin{eqnarray}
    & & \Mae I_{1}I_{n}j_{\lambda}(\vecvar{x};\omega,1/a)=
    E_{1}(\lambda)I_{n}j_{\lambda}(\vecvar{x};\omega,1/a),
    \label{eqn:verification1}\\
    & & \Mae I_{2}I_{n}j_{\lambda}(\vecvar{x};\omega,1/a)=
    E_{2}(\lambda)I_{n}j_{\lambda}(\vecvar{x};\omega,1/a).
    \label{eqn:verification2}
  \end{eqnarray}
Equations \myref{eqn:verification1}, \myref{eqn:verification2}
and \myref{eqn:Keep_triangularity} for $I_{n}j_{\lambda}$
are respectively the same as eqs. \myref{eqn:Hi-Jack_eigenfunction_1},
\myref{eqn:Hi-Jack_eigenfunction_2} and
\myref{eqn:Hi-Jack_triangularity} for the Hi-Jack polynomial
$j_{\lambda}$, which means $I_{n}j_{\lambda}$ satisfies
the definition of
the Hi-Jack polynomial except for
normalization.
Our definition of the Hi-Jack polynomial
uniquely specifies the Hi-Jack polynomial.
So we conclude
that $I_{n}j_{\lambda}$ must coincide with $j_{\lambda}$ up to
normalization.
Thus we confirm that the Hi-Jack polynomials $j_{\lambda}$
simultaneously diagonalize all the commuting conserved operators
$I_{n}$, $n=1,\cdots,N$. The eigenvalues of the conserved operators,
\[
  I_{n}j_{\lambda}(\vecvar{x};\omega,1/a)=E_{n}(\lambda)
  j_{\lambda}(\vecvar{x};\omega,1/a),
\]
are generally polynomials of the coupling parameter $a$:
\[
  E_{n}(a)=e_{n}^{(0)}(\lambda)+e_{n}^{(1)}(\lambda)a+\cdots.
\]
It is easy to get the constant ($a$-independent)
term $e_{n}^{(0)}(\lambda)$
because the term corresponds to the $n$-th eigenvalue for
$N$ non-interacting bosons confined in an external harmonic well:
\[
  e_{n}^{(0)}(\lambda)=\sum_{k=1}^{N}(\lambda_{k})^{n}.
\]
It is clear that there is no degeneracy in the constant terms of the
eigenvalues $\{e_{n}^{(0)}(\lambda)|n=1,\cdots,N\}$.
Since the conserved operators $I_{n}$ are Hermitian operators
concerning the inner product,
\[
  \langle j_{\lambda},j_{\mu} \rangle = \int_{-\infty}^{\infty}
  \prod_{k=1}^{N}{\rm d}x_{k}|\langle\vecvar{x}|0\rangle|^{2}
  j_{\lambda}j_{\mu} \propto \delta_{\lambda,\mu},
\]
the Hi-Jack polynomials are the orthogonal symmetric polynomials with
respect to the above inner product.
From the explicit form of the weight function,
\[
  |\langle\vecvar{x}|0\rangle|^{2}=
  \prod_{1\leq j<k\leq N}|x_{j}-x_{k}|^{2a}
  \exp\Bigl(-\omega\sum_{l=1}^{N}x_{l}^{2}\Bigr),
\]
we conclude that the Hi-Jack polynomial is a multivariable
generalization of the Hermite polynomial~\cite{Lasselle_1}.

\section*{Acknowledgements}\hspace*{\parindent}
One of the authors (MW) thanks Professor L. Vinet and workshop organizers
for their warm hospitality during the conference. The other author (HU)
appreciates Research Fellowships of the Japan Society for
the Promotion of Science for Young Scientists.


\begin{thebibliography}{UWH}
\bibitem[A]{Avan_5} J. Avan: Phys. Lett. A {\bf 185} (1994) 293.

\bibitem[AJ1]{Avan_2} J. Avan and A. Jevicki:
Phys. Lett. B {\bf 266} (1991) 35.

\bibitem[AJ2]{Avan_3} J. Avan and A. Jevicki:
Phys. Lett. B {\bf 272} (1991) 17.

\bibitem[AJ3]{Avan_4} J. Avan and A. Jevicki:
Commun. Math. Phys. {\bf 150} (1992) 149.

\bibitem[AT]{Avan_1} J. Avan and M. Talon: Phys. Lett. B {\bf 303} (1993) 33.

\bibitem[BR]{Barucchi_1} G. Barucchi and T. Regge:
J. Math. Phys. {\bf 18} (1977) 1149.

\bibitem[C]{Calogero_1} F. Calogero: J. Math. Phys. {\bf 12} (1971) 419.

\bibitem[D]{Dunkl_1} C. F. Dunkl:
Trans. Amer. Math. Soc. {\bf 311} (1989) 167.

\bibitem[J]{Jack_1} H. Jack:
Proc. R. Soc. Edinburgh (A) {\bf 69} (1970) 1.

\bibitem[La]{Lasselle_1} M. Lasselle:
C. R. Acad. Sci. Paris. t. S\'{e}ries I {\bf 313} (1991) 579.

\bibitem[LV]{Lapointe_1} L. Lapointe and L. Vinet: Commun. Math. Phys.
{\bf 178} (1996) 425.

\bibitem[M]{Moser_1} J. Moser: Adv. Math. {\bf 16} (1975) 197.

\bibitem[Pe]{Perelomov_1} A. M. Perelomov:
Theor. Math. Phys. {\bf 6} (1971) 263.

\bibitem[Po]{Polychronakos_1} A. P. Polychronakos:
Phys. Rev. Lett. {\bf 69} (1992) 703.

\bibitem[St]{Stanley_1} R. P. Stanley: Adv. Math. {\bf 77} (1988) 76.

\bibitem[Su1]{Sutherland_1} B. Sutherland: J. Math. Phys. {\bf 12} (1971) 246.

\bibitem[Su2]{Sutherland_2} B. Sutherland: Phys. Rev. A {\bf 4} (1971) 2019.

\bibitem[UHW]{Ujino_0} H. Ujino, K. Hikami and M. Wadati:
J. Phys. Soc. Jpn. {\bf 61} (1992) 3425.

\bibitem[UWH]{Ujino_1} H. Ujino, M. Wadati and K. Hikami:
J. Phys. Soc. Jpn. {\bf 62} (1993) 3035.

\bibitem[UW1]{Ujino_2} H. Ujino and M. Wadati: J. Phys. Soc. Jpn. {\bf 63}
(1994) 3585.

\bibitem[UW2]{Ujino_3} H. Ujino and M. Wadati: J. Phys. Soc. Jpn. {\bf 64}
(1995) 39.

\bibitem[UW3]{Ujino_4} H. Ujino and M. Wadati: J. Phys. Soc. Jpn. {\bf 64}
(1995) 2703.

\bibitem[UW4]{Ujino_6} H. Ujino and M. Wadati: J. Phys. Soc. Jpn. {\bf 65}
(1996) 653.

\bibitem[UW5]{Ujino_7} H. Ujino and M. Wadati: J. Phys. Soc. Jpn.
{\bf 65} (1996) 2423.

\bibitem[UW6]{Ujino_5} H. Ujino and M. Wadati: J. Phys. Soc. Jpn. {\bf 66}
(1997) 345.

\bibitem[U1]{Ujino_8} H. Ujino:
Orthogonal Symmetric Polynomials Associated with the Calogero Model,
to appear in Extended and Quantum Algebras and their Applications
in Mathematical Physics, proceedings of the Canada-China Meeting in
Mathematical Physics, Tianjin, China, August 19--24, 1996,
ed. M.-L. Ge, Y. Saint-Aubin and L. Vinet (Springer-Verlag).

\bibitem[U2]{Ujino_9} H. Ujino: Algebraic Study on the Quantum Calogero
Model, Doctoral Thesis, Univ. of Tokyo, December, 1996.

\end{thebibliography}
\end{document}